\documentclass[12pt,preprint]{aastex}
\newcommand{\Msun}{~M_\odot}

\newcommand{\msun}{M_\odot}

\newcommand{\kms}{\rm ~km~s^{-1}}

\newcommand{\ml}{\Msun ~\rm yr^{-1}}

\begin{document}

\title{COMMON ENVELOPE EVOLUTION LEADING TO  SUPERNOVAE WITH DENSE INTERACTION}
\author{Roger A. Chevalier}
\affil{Department of Astronomy, University of Virginia, P.O. Box 400325, \\
Charlottesville, VA 22904-4325; rac5x@virginia.edu}

\begin{abstract}
A variety of supernova events, including Type IIn supernovae and ultraluminous supernovae,
appear to have lost up to
solar masses of their envelopes in 10's to 100's of years leading up to
the explosion.
In order to explain the close timing of the mass loss and supernova events, we
explore the possibility that the mass loss is driven by common envelope evolution
of a compact object (neutron star or black hole) in the envelope of a massive star
and the supernova is triggered by the inspiral of the compact object to the central
core of the companion star.
The expected rate of such events is smaller than the observed rate of Type IIn supernovae
but the rates may agree within the uncertainties.
The mass loss velocity is related to the escape velocity from the common
envelope system and is
comparable to the observed velocity of 100's of km s$^{-1}$ in Type IIn events.
The mass loss is expected to be denser near the equatorial plane of the binary system
and there is good evidence that the circumstellar media in Type IIn supernovae
are asymmetric.
Some of these supernova types show evidence for energies in excess of the canonical
$10^{51}$ ergs, which might be the result of explosions  from rapid
accretion onto a compact object through a disk.

\end{abstract}


\section{INTRODUCTION}

There is growing evidence for some supernovae exploding into a medium
that is much denser than the stellar winds that might be expected around a normal 
star.
In Type IIn (narrow line) supernovae \citep{schlegel90},
the optical luminosities are plausibly explained as being due to circumstellar interaction
and the circumstellar density can be estimated from the luminosity
\citep{chugai94}.
If narrow line widths are indicative of the presupernova outflow velocities, the typical
outflow velocities are $100-500\kms$, leading to times of mass loss before
explosion of $10-300$ yr and mass loss rates of $0.02-0.1\ml$ 
for typical Type IIn supernovae (SNe IIn) \citep{kiewe12}.
The mass loss can be up to several $M_{\sun}$ extending out as far as $10^{17}$ cm.

The class of ultraluminous supernovae overlaps the SNe IIn, with objects
like SN 2006gy that was very bright for 240 days and radiated $\ga 2\times 10^{51}$ ergs
in optical light \citep{miller10}.
Another group of the ultraluminous events are not SN IIn,
but have spectra that resemble SNe Ic at later times \citep{quimby11,pastorello10}.
\cite{chevalier11} suggested that the ultraluminous supernovae are due to
dense circumstellar interaction, but only ones with a circumstellar extent
greater than the radius at which radiation can diffuse out have Type IIn
characteristics.
The mass loss involved can be $\ga 10\msun$ and extends to $\ga 2\times 10^{15}$ cm 
for Type IIn characteristics \citep[see also][]{smithmccray07}.

To account for such high mass loss rates, luminous blue variable (LBV) progenitors
have been suggested \citep[][and references therein]{smith10,kiewe12}.
In the case of SN 2005gl, a progenitor object was observed that is consistent with
an LBV \citep{galyam09}.
However, the LBV possibility does not answer the question 
of why is the explosion so well synchronized with the strong mass loss event;
the LBV phase is expected to be followed by a Wolf-Rayet phase lasting $10^5-10^6$ yr.
Another point is that LBVs are associated with very massive stars
($\sim 30-80\Msun$), but the stellar populations around SNe IIn  are comparable to those around
SNe II in general, which typically come from lower mass stars \citep{kelly11,anderson12}.
These results for the Type IIn are distinct from the  Types IIb and Ib/c which
show a clear connection to star forming regions.
The implication is that SN IIn progenitors are not confined to very high mass stars,
but may cover a broad range of stellar masses \citep{kelly11,anderson12}.

These properties argue against a particular
mass range becoming a supernova, and indicate that some factor other than mass
plays a role.
Here we suggest that the factor is 
binarity and that the mass loss and explosion
are both driven by the inspiral of a compact object in common envelope (CE)
evolution (Section 2).
This explosion mechanism has been previously considered by \cite{fryer98}, \cite{zhang01},
\cite{barkov11}, and \cite{thone11}.
The implications of the CE mass loss for Type IIn supernovae and related objects
are presented in Section 3.
General aspects of the mechanism are discussed in Section 4.

\section{COMMON ENVELOPE EVOLUTION AND EXPLOSION}

The suggested sequence of events leading to supernovae
with dense environments is shown in Figure 1.
The starting point is two massive stars in a binary.
The more massive star evolves, transfers mass to its companion, and
explodes as a supernova, leaving a neutron star (NS) or, less likely, a black hole (BH).
In the cases where the NS remains bound in a close orbit to the companion
star, the binary enters a common envelope (CE) phase when the companion evolves
and expands.
Depending on the initial separation, the CE phase starts only when the companion
becomes a red supergiant, or at an earlier phase if the binary is tighter.
During the inspiral phase, the Bondi-Hoyle accretion rate is well above the $10^{-3}\ml$
limit for spherical neutrino cooled accretion \citep{chevalier93}.
However, angular momentum of the
accreting material can prevent the high pressures needed for neutrino emission
 \citep{chevalier96} and numerical simulation of CE evolution show
 that the accretion rate can be significantly smaller than the
Bondi-Hoyle value \citep{ricker12}.
Whether neutrino cooled accretion can occur during spiral in through the envelope is
controversial \citep{fryer98,barkov11}, but the outcome is not crucial to the present
arguments.

The initial progress of the inspiral process depends on the density profile in the companion star
\citep{taam00}.
A red supergiant has a moderately flat density profile in the envelope, leading
to strong interaction followed by weak interaction and envelope ejection, and
leaving a NS/He star binary (Figure 1).
Because CE evolution with a red supergiant occurs over a relatively small range
of separations, it is more likely that the CE phase occurs at an earlier evolutionary
phase when there is a steady gradient from the center to the edge of the star,
and inspiral does not stop outside the stellar core.

A possible outcome of the inspiral to the central core is the formation of
a Thorne-Zytkow object \citep[TZO,][]{thorne77}: an  extended
star with a NS core.
TZOs have not been observationally identified, so their existence remains uncertain.
\cite{chevalier96} argued that inspiral in the core of the companion would lead to
rapid, neutrino-cooled accretion and collapse to a black hole because of the
high densities present there.
For inspiral in a He star, \citep{fryer98} estimated accretion rates as high
$3\times 10^7\ml$, leading not only to BH formation but also strong explosions.
SPH simulations of the process by \cite{zhang01} showed accretion rates
of $10^5 - 10^6\ml$, sufficient for a $10^{51}$ erg explosion.
In the present scenario,  the interaction of such an explosion with the mass lost during the initial CE
phase is potentially a Type IIn supernova.
Figure 1 shows the TZO and SN outcomes of central inspiral; additional possible outcomes
are BH without a supernova or a magnetar that was spinning rapidly enough to
eject surrounding material.

The expansion velocity that is expected from common envelope evolution is
of order the escape velocity from the extended star \citep{terman95, taam00}.
For a red supergiant companion, this is $\la 100\kms$ which is low
for a Type IIn circumstellar medium, but this evolution channel
is expected to lead to a He star/compact object binary and not directly to a supernova.
The lifetime of the He star is $\la 10^5$ yr, so that the radius of the H envelope
at the time that the He star reaches its advanced burning phases is $\sim 10^{19}$ cm.
For CE evolution at an earlier evolutionary phase, the star is more compact and
the escape velocity is higher.  
For example, in case 1 of \cite{terman95}, the $16\Msun$ star has a
radius of $2\times 10^{12}$ cm and the escape
velocity is $430\kms$.
It is the more compact cases in which the compact object can continue to spiral in
to the core and give rise to a supernova in the present scenario.

Another factor for determining the disposition of the circumstellar matter at the time
of the explosion is the time between the beginning of mass loss and the explosion.
The explosion depends on the rapid accretion of matter onto the
black hole, which occurs only when the BH has reached the central region of the stellar core,
although scenarios with slower accretion leading to a magnetar may also
be possible \citep{barkov11}.
The timescale for inspiral to occur is $\la 10^3$ yr \citep{taam00}.
The timescale for the initial orbit decay is $\sim 200$ yr for a star in its early core He
burning phase, but $\sim 3$ yr for a star in its late core He burning phase, because of
the difference in the density distribution \citep{terman95,taam00}.

An expectation for mass lost during the common envelope phase is that it is
concentrated toward the orbital plane of the binary \citep{terman95,taam10}.
In 3-dimensional simulations involving a $1.05\Msun$ red giant and a lower mass companion,
\cite{ricker12}
find that 90\%  of the mass outflow is in a region 30$^{\circ}$ about the orbital plane.
Figure 7 of \cite{ricker12} indicates that the mass per unit solid angle in the equatorial
direction is 20 times that in the polar direction.
The degree of asymmetry depends on the mass ratio (secondary to primary), with a small ratio giving a
higher degree of asymmetry.
In the case of a  ratio near unity, the stellar core is spun up by the companion, leading to a broad
flow with an evacuated region along the poles \citep{taam10}.
In the case considered here, the mass ratio is fairly small.

For systems that form a stable He star/neutron star binary,
the further possibilities for the binary are inspiral into the He star, or the explosion of the
He star leading to two compact objects either bound or unbound (Figure 1).
In the present scenario, the channel that leads to inspiral in the  He star can give rise
to H-free luminous supernovae \citep{quimby11,pastorello10}
and possibly Type Ibn supernovae \citep{pastorello08} or low luminosity
gamma-ray bursts (GRBs) \citep{thone11}.

\section{COMPARISON WITH TYPE IIn AND RELATED SUPERNOVAE}

The basic properties of SNe IIn surroundings given in Section 1 (velocity, mass,
and timescale) are in rough agreement with mass loss during a CE phase.
The extent of the mass loss is important for the supernova properties.
If the extent is relatively small ($\sim 1\times 10^{16}$) cm, the supernova is very luminous for 
a relatively short time ($\sim 240$ days as for SN 2006gy \citep{miller10}),
while if the extent is large ($\sim 1\times 10^{17}$) cm, the supernova is less luminous
but for a long time (10 yr as for SN 1988Z \citep{aret99}).
In both of these cases, the observed radiated energy is $\ga (1-2)\times 10^{51}$ erg.

A further consequence of the CE scenario for the formation of the circumstellar medium is that it
should have higher density in the equatorial plane.
The possibility that SN IIn involve interaction with equatorial mass loss was already
suggested by \cite{chugai94} to explain the intermediate velocity ($\sim 2000\kms$)
component in spectra of SN 1988Z.
The presence of both high velocity and intermediate velocity components required either
a clumpy circumstellar medium or an asymmetric flow.
\cite{chugai94} preferred the clumpy wind model because the mass loss rate required by
the equatorial flow model ($0.015\ml$) was considered too high.
However, this rate is in keeping with expectations for a CE phase of evolution.
The most direct evidence for asymmetric interaction comes from polarization measurements.
High  polarization at optical wavelengths has been observed in the SNe IIn that have been studied
\citep[][and references therein]{patat11}.
The continuum polarization is likely due to electron scattering in an asymmetric circumstellar
medium, suggesting an axial ratio $\la 70$\% in the case of SN 2010jl \citep{patat11}.

\cite{chugai94} note that a distinction between the clumpy and equatorial structure is
that the intermediate component
emission is near the outer shock in the clumpy case, but is at a relatively small
radius (corresponding to the velocity) in the equatorial case.
It has not been possible to spatially resolve the intermediate component 
optical emission in any SN IIn, but it has been
possible to resolve the radio emission from the nearby Type IIn SN 1986J with VLBI
techniques \citep{bieten10}.
In the case of SN 1986J, the width 
of the ``intermediate width'' H$\alpha$ line was $\sim 1000\kms$ (FWHM) at an age of 3 yr
\citep{rupen87} and remained narrow at 24 yr \citep{mili08}.
Recent radio observations show a centrally located source that is marginally resolved
at $5\times 10^{16}$ cm \citep{bieten10}, corresponding to an average radial velocity 
of $340\kms$ at an age of 23 yr.
It is possible that the central emission is associated with the inner equatorial
interaction region.

To estimate the rate of the mergers of NSs with stars, we use previous estimates of the rate of formation
of TZOs because we are following the hypothesis that evolution leading to these objects
can instead lead to supernovae.
Based on numbers of high mass X-ray binaries,
\cite{podsiad95} estimated a rate of TZO formation in the Galaxy of $\ga 2\times 10^{-4}$ yr$^{-1}$, 
which includes inspiral events and mergers due to the direction of the neutron star kick
during formation.
The core collapse supernova (CCSN) rate in the Galaxy is $ (2.30\pm 0.48)\times 10^{-2}$ yr$^{-1}$ 
and the SN II rate is $ (1.52\pm 0.32)\times 10^{-2}$ yr$^{-1}$    \citep{li11b}.
\cite{li11a} find that $8.6^{+3.3}_{-3.2}$\% of SN II are of Type IIn, leading to a SN IIn rate
of $13\times 10^{-4}$ yr$^{-1}$, while \cite{smartt09} find that 3.8\% of CCSN are IIn,
leading to a IIn rate of $8.7\times 10^{-4}$ yr$^{-1}$.
The Type IIn rate in the Galaxy may be somewhat decreased because IIn events preferentially
occur in low mass galaxies \citep{li11a}.
The result is that the merger rate falls short of the SN IIn rate, but the substantial uncertainty
in the rates might allow the events to be associated.

As noted above, the rate of inspirals stopping outside the core is expected to be smaller
than the rate of inspirals continuing into the core (Figure 1).
Binary NS or BH systems should thus form less frequently than central inspirals.
The rate of formation of binary compact objects (NS or BH) by the channel
shown in Figure 1 is estimated at
$\sim 0.9\times 10^{-4}$ yr$^{-1}$ in binary
population models \citep{bel02}.

\section{DISCUSSION}

There is good evidence that the most luminous SNe IIn occur in low
metallicity regions \citep{neill11}.
In the scenario suggested here, this property can be attributed to the dependence
of mass loss on metallicity.
At solar metallicity, stars with masses above $\sim 35\Msun$ do not become red
supergiants, but remain relatively compact because of mass loss.
At low metallicity, this mass limit is expected to increase and more massive stars become
extended.
The expansion in radius is needed to enter the CE phase, so more 
massive stars can experience spiral in and explosion at low metallicity.
These stars have more mass loss and more massive black holes, so that more
luminous supernovae are expected.

The energies of well-observed Type II supernovae, including SN 1987A and SN 1993J,
are  $\sim 1\times 10^{51}$ ergs,
while high energies, $\ga 10^{52}$ ergs, have been inferred for some Type Ic supernovae
that are associated with GRBs \citep[e.g., Figure 8 in][]{tanaka09}.
In the GRB case, the high supernova energy is presumably 
associated with the central engine, but
the exact mechanism for the high energy is not understood.
High energies have also been inferred for SNe IIn primarily from their radiated energy,
 e.g., SN 1998Z is estimated to have had 
 a total radiated energy as high as  $5\times 10^{51}$ \citep{aret99} and the explosion
energy must have been significantly higher.
A high explosion energy does not appear to be due to a highly massive star progenitor;
 the interaction properties of SN 1988Z indicate an ejecta mass
$\la 1\Msun$ \citep{chugai94}.
The link between central engines and energetic supernovae may extend to SNe IIn.

A clearer connection to a rapidly rotating central engine would be evidence for
a jet-like flow in a SN IIn.
Although there have been suggested associations of SNe IIn with  GRBs, e.g. SN 1997cy
 \citep{germany00},
there have been no convincing associations, which can be attributed to there being
too much surrounding dense material  for a relativistic flow to propagate.
However, the Type IIn SN 2010jp shows evidence for a nonrelativistic jet-like flow
\citep{smith12}.

The Type IIn phenomenon has also been observed in association with Type Ia supernovae,
starting with SN 2002ic \citep{hamuy03}.
The scenario of CE mass loss followed by a supernova can also be considered for
the white dwarf case.
A white dwarf spirals into the envelope of an evolved companion and continues to
the core where strong accretion gives rise to a thermonuclear explosion.
This scenario would be compatible with a double degenerate origin for Type Ia
supernovae, as discussed by \cite{livio03}.

The proposal made here about the origin of the dense matter around Type IIn and related
supernovae is speculative, but shows some promising points of comparison.
The possibility that CE evolution leads to the matter has long been mentioned
\citep{chugai94}. 
The hypothesis that the mass loss and
supernova both result from CE evolution has been raised by \cite{barkov11} and (after this
paper was submitted) by \cite{soker12}, but has not been much explored.
There are other suggestions for the presence of the dense matter near a SN IIn.
\cite{metzger10} proposed disk material that has survived from the however formation phase;
however, the narrow P Cygni lines that are observed in SNe IIn are generally taken
to imply outflowing dense material, and there is no direct evidence for such disks around
massive stars.
\cite{quataert12} proposed mass loss driven by gravity waves generated in the late burning phases;
however, at this point the model does not explain what are the properties of the small fraction
of massive stars that end as SN IIn events.
The binary model proposed here provides a natural explanation for the heterogeneity
of Type IIn events.
For the case of a very massive star, it explains the presence of H at the time
of the supernova, which would not be present if the star ran its full evolutionary course.
Whether the binary hypothesis is viable ultimately depends on a better understanding of the
CE process.
The process is complex, but detailed, 3-dimensional simulation are becoming
feasible \citep{ricker12}.

\acknowledgments
I thank  Mel Davies, Mario Livio, and Ron Taam  for discussions, and the referee for helpful comments.
This research was supported in part by NSF grant  AST-0807727.


\clearpage

\begin{figure}[!hbtp]   
\epsscale{1.0}
\plotone{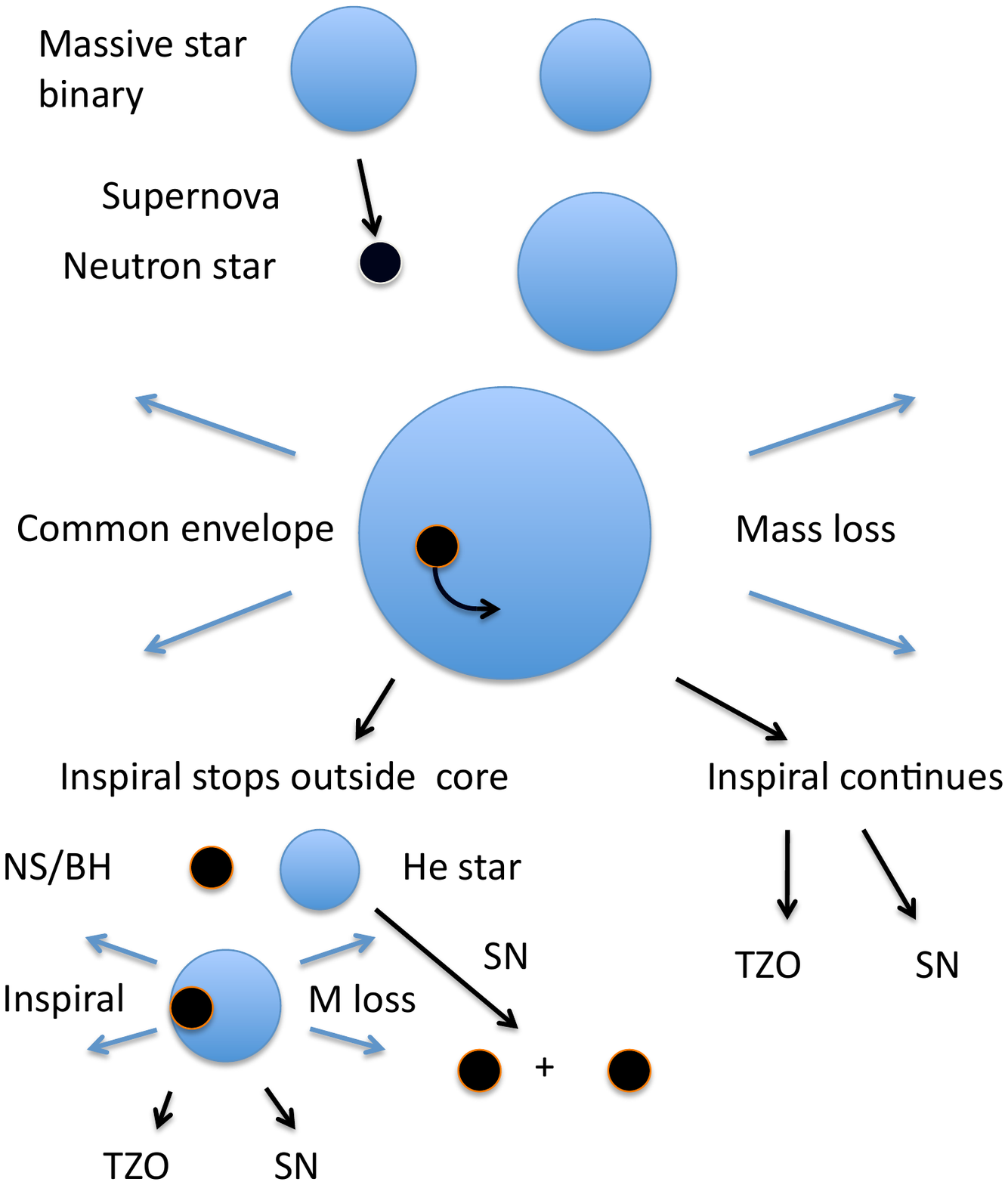}
\caption{The binary evolution possibly leading to supernovae in a dense mass loss region.
Here NS is neutron star, BH is black hole, TZO is Thorne-Zytkow object, and SN is supernova.
The supernovae of interest here are to the bottom right, where H is present, and to the
bottom left, where H is absent.
}
\end{figure}

\end{document}